\documentclass[10pt,conference]{IEEEtran}
\IEEEoverridecommandlockouts
\usepackage{cite}
\usepackage{amsmath,amssymb,amsfonts}
\usepackage{algorithmic}
\usepackage{graphicx}
\usepackage{textcomp}
\usepackage{xcolor}
\usepackage{orcidlink}
\usepackage{eso-pic}

\usepackage[T1]{fontenc}
\usepackage{booktabs}
\usepackage{multirow}
\usepackage{makecell}
\usepackage{pifont}

\def\BibTeX{{\rm B\kern-.05em{\sc i\kern-.025em b}\kern-.08em
    T\kern-.1667em\lower.7ex\hbox{E}\kern-.125emX}}
\begin{document}

\title{FeatureFox: Sample-Efficient Panoptic Graph Segmentation for Machining Feature Recognition in B-Rep 3D-CAD Models}

\author{
\IEEEauthorblockN{
Bertram Fuchs\textsuperscript{1,2}\,\orcidlink{0009-0006-5107-428X}\quad
Altay Kacan\textsuperscript{2,3}\,\orcidlink{0000-0003-3438-2281}\quad
Aaron Haag\textsuperscript{2}\,\orcidlink{0009-0007-0967-9354}\quad
Oliver Lohse\textsuperscript{2}\,\orcidlink{0000-0001-8450-2398}
}
\IEEEauthorblockA{
\textsuperscript{1}Technische Universit\"at Berlin\quad
\textsuperscript{2}Siemens AG, M\"unchen, Germany\\
\textsuperscript{3}Friedrich-Alexander-Universit\"at Erlangen-N\"urnberg\\
\{bertram.fuchs, altay.kacan, aaron.haag, oliver.lohse\}@siemens.com
}
}

\maketitle

\AddToShipoutPictureFG*{%
  \AtPageLowerLeft{%
    \hspace{0.5in}%
    \raisebox{0.25in}{%
      \parbox{\dimexpr\paperwidth-1in\relax}{%
        \centering\scriptsize
        This work has been submitted to the IEEE for possible publication. Copyright may be transferred without notice, after which this version may no longer be accessible.
      }%
    }%
  }%
}

\begin{abstract}
Automatic feature recognition (AFR) on B-Rep 3D-CAD models is central to CAD/CAM automation, yet most learning-based methods are complex, data-hungry, and evaluate instance grouping and semantic labeling separately. We present FeatureFox, a \emph{panoptic AFR} pipeline that outputs machining \emph{instances with semantic labels}: a calibrated binary edge classifier on enriched edge attributes localizes feature boundaries, instances are recovered as connected components in a pruned face-adjacency graph, and a per-instance classifier predicts the machining class from aggregated subgraph attributes. We evaluate on MFInstSeg using Panoptic Quality (PQ), which jointly scores instance separation and semantic correctness. FeatureFox is substantially more sample- and compute-efficient than the deep baseline AAGNet, reaching $\mathrm{PQ}>0.9$ with $\sim250$ training parts versus $\sim5{,}000$ for AAGNet, and training on the full MFInstSeg set takes seconds on a GPU. On the full training set, AAGNet surpasses FeatureFox marginally in PQ, while FeatureFox remains slightly ahead in feature-level recognition and localization accuracy. Finally, leveraging its low data requirement, we train FeatureFox on $270$ manually labeled industrial CAD parts and show qualitative generalization to an unseen real industrial part, indicating practical real-world applicability.
\end{abstract}


\section{Introduction}

Computer-aided process planning (CAPP) connects design to manufacturing by translating CAD models into CAM decisions such as setups, toolpaths, and operation sequences. One key bottleneck in this pipeline is automatic feature recognition (AFR): identifying manufacturable machining feature instances from CAD geometry and topology. When AFR is unreliable, downstream planning can become error-prone and may require manual intervention, limiting CAD/CAM automation.

AFR is challenging because real mechanical parts rarely present features in clean isolation. Features intersect, blend, and appear partially due to design variations or modeling conventions, and small geometric edits can significantly alter local topology. In addition, process planning requires more than semantic category labels: it requires separate feature instances with consistent boundaries. These requirements naturally align with the task of panoptic segmentation, where semantic and instance information must be inferred jointly.

Classic AFR approaches are knowledge-driven (rule-based systems, graph matching, volume decomposition, heuristics) and remain attractive due to interpretability \cite{subrahmanyam1995overview,verma2010review}, but they are not robust under intersections as well as design variability and are costly to maintain when part complexity increases \cite{babic2008review,verma2010review}. Learning-based methods improve generalization and achieve strong results on B-Rep representations, yet state-of-the-art models are often complex, data-hungry, and typically evaluate instance grouping and semantic labeling as separate tasks. This creates a practical gap: labeled industrial CAD data is scarce and expensive, so methods that require thousands of annotated parts are not deployable for AFR in industry.

%Contributions
\vspace{0.5em}
\noindent\textbf{Contributions.} To address this, we present FeatureFox and make the following contributions:
\begin{itemize}
  \item[$\bullet$] \textbf{Panoptic AFR formulation and metric.} We cast machining feature recognition in 3D-CAD on B-Rep faces as a \emph{Panoptic Segmentation} problem (faces per instance + semantic instance label) and advocate \emph{Panoptic Quality (PQ)} as a single metric that jointly evaluates machining feature localization and class recognition.

  \item[$\bullet$] \textbf{FeatureFox: interpretable hierarchical pipeline.} We introduce FeatureFox, a lightweight tree-based panoptic AFR pipeline that (i) localizes feature instances via a calibrated binary edge-boundary model on enriched edge attributes, (ii) recovers instances as connected components in a pruned face-adjacency graph, and (iii) assigns one semantic label per \emph{whole instance} using aggregated subgraph attributes.

  \item[$\bullet$] \textbf{Strong sample- and compute-efficiency on MFInstSeg.} On AAGNet's MFInstSeg dataset, FeatureFox reaches $\mathrm{PQ}>0.9$ with $\sim250$ training parts, whereas the deep baseline AAGNet requires $\sim5{,}000$; training FeatureFox on the full MFInstSeg training set takes seconds on a GPU, potentially enabling offline customer-centric deployment.

  \item[$\bullet$] \textbf{Practical low-data industrial applicability.} Enabled by its low data requirement, we train FeatureFox on $\sim270$ manually labeled real-world CAD parts and demonstrate qualitative generalization to an unseen industrial part, indicating feasibility for real deployment scenarios with scarce annotations.
\end{itemize}

Due to internal company policy and confidentiality constraints, we cannot publicly release the code and data for this work.

\section{Related Work}

AFR has been researched since the 1980s \cite{shi2020critical} as one of the core problems to be solved in CAPP and broadly falls into rule-based and learning-based methods. Rule-based systems recognize machining features via handcrafted geometrical and topological rules or graph-based matching procedures, and remain attractive due to interpretability, but they often break on intersecting or topologically complex features and require substantial engineering effort to maintain at scale \cite{babic2008review,rahmani2007hybrid,shi2020manufacturing}.

Learning-based AFR progressed mainly by changing the input representation and moved toward CAD-native models. Early deep methods used voxels \cite{zhang2018featurenet}, multi-view renders \cite{shi2020novel,shi2020intersecting} or point-clouds \cite{lei2022mfpointnet}, but these representations either lose topological information or are computationally heavy and depend on view selection \cite{colligan2022hierarchical,shi2020novel}. More recent work operates directly on B-Rep–derived graphs \cite{colligan2022hierarchical} and combines geometric UV-grid sampling with message passing in a GNN \cite{jayaraman2021uv}. These models typically predict per-face semantic labels, i.e., semantic segmentation on B-Reps.

Only a few methods explicitly address instance-aware AFR, separating distinct feature instances in addition to semantic labeling. ASIN jointly clusters instances and predicts semantics from point clouds \cite{asin_zhang2022machining}, while AAGNet introduces a B-Rep–native gAAG representation with a multi-head CNN+GNN pipeline for semantic and instance predictions \cite{aagnet_wu2024aagnet}. MFTReNet follows AAGNet’s gAAG-based multi-task GNN design, but extends it with additional heads to predict inter-feature topological relations \cite{mftrenet_xia2024machining}. EAGIS targets instance grouping via edge relations on MFInstSeg but does not address semantic labeling \cite{li2025edge}. Overall, most prior learning-based methods either focus on semantic segmentation or instance segmentation only or do not provide a single unified panoptic output. To the best of our knowledge, all current State-of-the-Art learning-based AFR methods require thousands of labeled CAD parts and have not demonstrated training and evaluation on real industrial CAD files, limiting industrial applicability because annotations are scarce and costly.

\section{Methodology}

Previous methods that perform well on the task of AFR all train a hybrid model consisting of a CNN and a Graph Neural Network \cite{aagnet_wu2024aagnet,mftrenet_xia2024machining}. We define the task as a Panoptic Segmentation \cite{kirillov2019panoptic} task on Graphs and propose FeatureFox, a hierarchical calibrated Tree-based model that recognizes feature instances and their respective semantic class.

\subsection{Graph representation of B-Reps and features}

Modern CAD systems usually represent the topology and geometry of CAD parts using Boundary Representation (B-Rep). A B-Rep describes a solid by its enclosing surfaces, edges, and vertices, and their connectivity (topology) and geometry. As B-Rep is a variable-sized, CAD-kernel dependent geometric and topological data structure and not a fixed-dimensional numeric representation suitable for Machine Learning, a representation format called Geometric Attributed Adjacency Graph (gAAG) \cite{aagnet_wu2024aagnet} was proposed. EAGIS changed the representation format and used edges as nodes for training a GNN, and formulated the problem of feature instance detection as binary edge prediction task \cite{li2025edge}. FeatureFox builds on the gAAG representation, and leverages face adjacency to propagate neighboring face information to edges.

\paragraph{Face-adjacency graph.}
Given a B-Rep with faces $\mathcal{F}=\{f_1,\dots,f_n\}$ and B-Rep edges $\mathcal{E}=\{e_1,\dots,e_m\}$, let
$\mathrm{Inc}:\mathcal{E}\rightarrow 2^{\mathcal{F}}$ map each B-Rep edge to its incident faces.
Assuming a 2-manifold B-Rep, each edge is shared by exactly two faces:
\begin{equation}
\forall e\in\mathcal{E}:\quad \mathrm{Inc}(e)=\{f_i,f_j\},\;\; i\neq j .
\end{equation}
We define the undirected face-adjacency graph $G=(V,E)$ with one node per face,
\begin{equation}
V := \mathcal{F},
\end{equation}
and an (undirected) graph edge between two nodes iff the corresponding faces share a B-Rep edge:
\begin{equation}
E := \big\{\{f_i,f_j\}\subseteq\mathcal{F}\;:\;\exists e\in\mathcal{E}\ \text{s.t.}\ \mathrm{Inc}(e)=\{f_i,f_j\}\big\}.
\end{equation}
Equivalently, the adjacency matrix $A\in\{0,1\}^{n\times n}$ is
\begin{equation}
A_{ij}=
\begin{cases}
1, & \exists e\in\mathcal{E}: \mathrm{Inc}(e)=\{f_i,f_j\},\ i\neq j,\\
0, & \text{otherwise}.
\end{cases}
\end{equation}

\paragraph{Per-face and per-edge attributes.}
Let the B-Rep contain $n = |\mathcal{F}|$ faces and $m = |\mathcal{E}|$ (manifold) edges, and let
$\mathrm{Inc}:\mathcal{E}\to 2^{\mathcal{F}}$ map each B-Rep edge to its incident faces with
$\mathrm{Inc}(e_k)=\{f_{s(k)}, f_{t(k)}\}$, where $s,t:\{1,\dots,m\}\to\{1,\dots,n\}$ assign an arbitrary (but fixed) ordering of the two incident face indices of edge $e_k$.
We associate one attribute vector to each face and to each B-Rep edge, where $d_F$ is the dimensionality of the face attribute vector:
\begin{equation}
\mathbf{x}^{F}_i \in \mathbb{R}^{d_F}\quad (i=1,\dots,n), 
\qquad
\mathbf{x}^{E}_k \in \mathbb{R}^{d_E}\quad (k=1,\dots,m),
\end{equation}.

\paragraph{Face UV-grid tensor.}
For each face $f_i$ we sample a regular grid like in UV-Net \cite{jayaraman2021uv} in its parametric domain with resolution
$U\times V$. For grid indices $(u,v)$ we store 3D position $\mathbf{p}_i(u,v)\in\mathbb{R}^3$,
unit normal $\mathbf{n}_i(u,v)\in\mathbb{R}^3$, and a binary validity mask $\mu_i(u,v)\in\{0,1\}$
indicating whether the sample lies on the trimmed face.
We stack these three in the face-grid tensor $\mathbf{T}^{F}\in \mathbb{R}^{n\times U\times V\times 7}$.

\paragraph{Edge sampling tensor.}
Analogously, we sample each B-Rep edge $e_k$ with $M$ points along its parameter and store a
15-dimensional geometric descriptor per sample. This yields the edge tensor $\mathbf{T}^{E}\in \mathbb{R}^{m\times M\times 15}$ that stacks for each edge sample the 3D point,
$\mathbf{t}_k(j)\in\mathbb{R}^3$, the unit tangent, $\mathbf{n}_k(j)\in\mathbb{R}^3$ the unit normal,
$\mathbf{b}_k(j)\in\mathbb{R}^3$ the unit binormal, $\mu_k(j)\in\{0,1\}$ a trim mask indicating whether
the sample lies inside or on the valid edge, $r_k(j)\in\mathbb{R}$ the (local) radius, and
$\tau_k(j)\in\mathbb{R}$ the torsion.

\subsection{Instance Segmentation Model}

For training our instance model, we first enrich our edge attributes with additional information from neighboring nodes and then train a calibrated binary feature boundary classifier. 

\paragraph{Enriched edge attributes}

\begin{figure}[t]
\centering
\includegraphics[width=0.65\columnwidth]{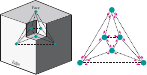}
\caption{Graphical representation and information propagation from vertices (B-Rep faces) to edges (B-Rep edges). Face Adjacency Graph as used in UV-Net \cite{jayaraman2021uv} (Left). Edge samples and per-edge attributes are used to calculate the per-edge attribute feature vectors. Per-face attributes and additional attributes calculated from per-face samples are propagated towards the edges and enrich the edge attributes (Right).}
\label{fig:featfox_graph_representation}
\end{figure}

For each graph edge $k\in E$ (corresponding to a B-Rep edge incident to exactly two faces), let $i$ and $j$ denote the indices of its two incident faces in the face-adjacency graph; we refer to this edge by the unordered pair $\{i,j\}$. 
We define additional hand-crafted edge features by combining information from the observed edge and its two neighboring faces, and concatenate these features to the pre-extracted edge attribute vector $\mathbf{x}^{E}_k$. Concretely,
\[
\tilde{\mathbf{x}}^{E}_k \;=\; \big[\,\mathbf{x}^{E}_k \ \Vert\ \Delta\mathbf{x}^{E}_k\,\big],
\]
where $[\cdot\Vert\cdot]$ denotes concatenation. The information flow is depicted in figure \ref{fig:featfox_graph_representation}.

Edge attributes include the dihedral angle, concavity, area and perimeter ratios, normalized length and distance between face centroids and the edge class among others.

\paragraph{Labelling paradigm}

\begin{figure}[t]
\centering
\includegraphics[width=0.65\columnwidth]{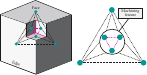}
\caption{FeatureFox binary machining feature instance labels. We are labeling according to the question: Are both faces adjacent to the edge in the same machining feature instance? Edges representing machining feature boundaries are labeled with False (black), edges within one feature instance are labeled with True (pink).}
\label{fig:featfox_binary_feature_labels}
\end{figure}

We formulate feature instance localization as a binary edge-classification problem. 
For each graph edge $k\in E$ connecting two adjacent faces with indices $i$ and $j$, let $\ell(i)\in\{1,\dots,L\}$ denote the ground-truth machining feature instance ID of face $i$, where $L$ is the number of feature instances. 
We then define the binary edge label

\[
y_k \;=\; \left[\ell(i)=\ell(j)\right] \;\in\;\{0,1\},
\]
i.e., an edge is labeled positive iff both incident faces belong to the same machining feature instance. 
Figure~\ref{fig:featfox_binary_feature_labels} illustrates this labeling: $y_k=1$ (True, pink) if $\ell(i)=\ell(j)$, and $y_k=0$ (False, black) otherwise.

\paragraph{Per-Edge Calibrated Feature Instance Segmentation}

We train a calibrated binary \texttt{XGBoost} classifier with $200$ trees (maximum depth $6$, learning rate $0.1$) on the enriched edge features $\tilde{\mathbf{x}}^{E}_k$ to predict the probability $p(y_k=1\mid \tilde{\mathbf{x}}^{E}_k)$ that the two incident faces of edge $k$ belong to the same machining feature instance. Since our instance localization relies on probability thresholding to decide whether an edge should connect faces within the same instance, well-calibrated probabilities are crucial for a stable and interpretable decision threshold. We therefore apply isotonic calibration using three-fold cross-validation, which is sufficient in our setting because the calibration folds contain a large number of edge samples as each sample B-Rep contains many edges. Finally, we apply a fixed threshold to the calibrated probabilities to obtain binary feature-boundary predictions. We do not use the MFInstSeg validation set for model selection, since we fix all hyperparameters a priori and do not tune them as the resulting instance grouping performance is already high, as shown in Table~\ref{tab:mfinstseg_perf}.

We then recover machining feature instances as the connected components of a pruned face-adjacency graph. 
Specifically, given the original face-adjacency graph $G$ and predicted binary edge labels $\hat{y}_k\in\{0,1\}$, we retain only edges between faces in one feature, dropping feature boundary edges, and define the pruned graph $\hat{G}=(V,\hat{E})$ with
\[
\hat{E} \;=\; \{\, k\in E \mid \hat{y}_k = 1 \,\}.
\]
Each machining feature instance is then defined as a connected component of $\hat{G}$, i.e., a maximal set of faces that are mutually reachable via paths consisting only of edges in $\hat{E}$.

\paragraph{Per-instance Semantic Segmentation.}
\label{par:per_instance_semseg}

AAGNet performs semantic classification on a per-face level~\cite{aagnet_wu2024aagnet}. In contrast, we classify each \emph{feature instance} as a whole. This choice makes the pipeline explicitly panoptic: instance grouping is performed first, and semantics are then assigned per instance. For each ground-truth feature instance in $G$, represented by a connected \emph{node set}
$C\subseteq V$ (and its induced subgraph $G[C]$), we compute a single instance attribute vector $\mathbf{z}_C$ by aggregating the face sample, edge sample as well as face attribute and edge attribute vectors of all faces and edges contained in $C$, which can be compared to a manual form of pooling. Feature instance attributes include the number of faces and edges, the log characteristic length (used as scale descriptor)
\[
L_{\log}=\log\!\left(L+\varepsilon\right),\qquad 
L=\sqrt{\sum_{v\in C} A_v},
\]
where $C$ denotes the set of faces in the feature instance, $A_v$ is the area of face $v$, and $\varepsilon$ is a small constant for numerical stability, as well as a histogram of face and edge types, curvature distributions, graph topology features (e.g., spectral connectivity), and boundary-structure cues such as the number of disconnected boundary-edge loops, among others.
 We then train a semantic \texttt{XGBoost} classifier ($400$ trees, maximum depth $6$) on instance-label pairs $(\mathbf{z}_C, c_C)$, where $c_C$ is the ground-truth semantic label of instance $C$. As in the edge model, we fix all \texttt{XGBoost} hyperparameters a priori and therefore do not perform validation-based model selection. During inference, for each predicted instance subgraph $\hat{C}$ returned by the instance segmentation stage, we compute $\mathbf{z}_{\hat{C}}$ and output its semantic class prediction.

\begin{figure*}[!t]
\centering
\includegraphics[width=0.76\textwidth]{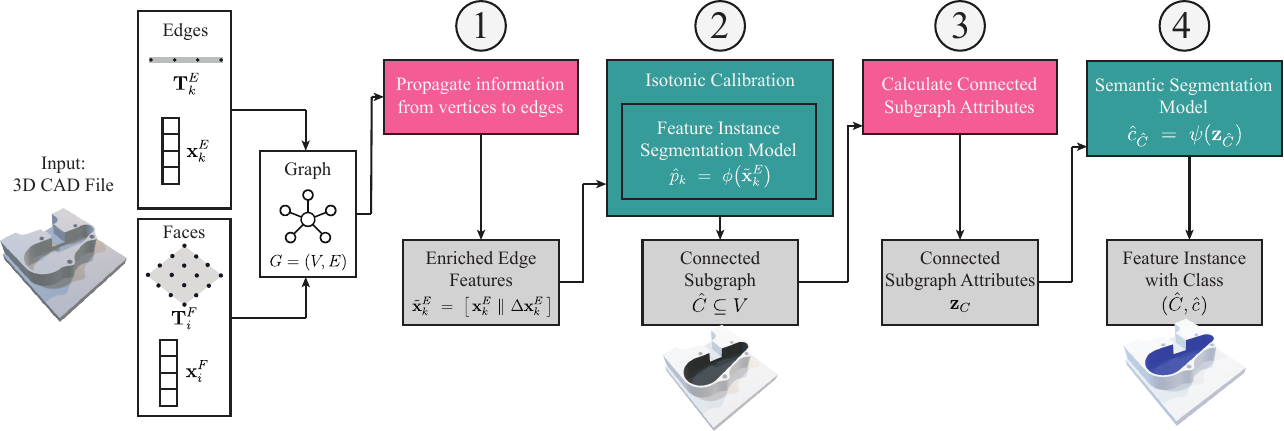}
\caption{FeatureFox pipeline. Starting from a B-Rep 3D-CAD model, we construct a face-adjacency graph and attributes, propagate vertex information to edges (1), and localize the feature with a calibrated edge boundary classifier for instance grouping (2). We then compute connected-component attributes (3) and recognize the class of each predicted instance with a semantic model (4). Pink: Attribute calculation. Teal: Model training. Gray: Intermediate representation.}

\label{fig:featfox_pipeline}
\end{figure*}

\section{Results}

We compare our algorithm on the dataset MFInstSeg, which was published by the authors of AAGNet, as it is the most frequently used AFR dataset and includes semantic and instance labels \cite{aagnet_wu2024aagnet,mftrenet_xia2024machining,dai2025brepformer,li2025edge}.

\subsection{Hardware Setup}

For extracting face and edge samples as well as face and edge attributes faster, we convert all 62477 STEP files (standardized CAD format) in MFInstSeg to Siemens NX .prt files. For all experiments we used 62178 MFInstSeg samples, for which the conversion was successful and no CAD topology or geometry integrity error was detected. For training AAGNet and FeatureFox we used an NVIDIA RTX 6000 Ada. 

\subsection{Face-level performance on MFInstSeg}

In a first ablation study, we evaluate instance segmentation performance of FeatureFox. Table \ref{tab:mfinstseg_perf} shows the instance segmentation performance in comparison to AAGNet and MFTReNet \cite{aagnet_wu2024aagnet,mftrenet_xia2024machining}. For FeatureFox, we do not state per-face semantic segmentation accuracy, as we train our semantic segmentation model for feature instance subgraphs and not on face-level as explained (see Methods section). FeatureFox performs similarly to AAGNet and MFTReNet for instance segmentation.

\begin{table}[t]
\centering
\caption{Per-Face performance on MFInstSeg test set.}
\label{tab:mfinstseg_perf}
\renewcommand{\arraystretch}{1.15}
\setlength{\tabcolsep}{6pt}

\resizebox{\columnwidth}{!}{%
\begin{tabular}{lccccc}
\toprule
\multirow{2}{*}{Network} &
\multicolumn{2}{c}{Semantic Segmentation} &
\multicolumn{2}{c}{Instance Grouping} &
\multirow{2}{*}{\makecell{Number of\\parameters  $\downarrow$}} \\
\cmidrule(lr){2-3}\cmidrule(lr){4-5}
& Acc(\%) $\uparrow$ & mIOU(\%) $\uparrow$ & Acc(\%) $\uparrow$ & $F_1$ Score $\uparrow$ & \\
\midrule
AAGNet &
\makecell{99.15\\$\pm$0.03} &
\makecell{\textbf{98.45}\\$\pm$0.04} &
\makecell{99.94\\$\pm$0.01} &
\makecell{98.84\\$\pm$0.07} &
\textbf{0.41 M} \\
MFTReNet &
\makecell{\textbf{99.56}\\$\pm$0.02} &
\makecell{98.43\\$\pm$0.03} &
\makecell{\textbf{99.95}\\$\pm$0.01} &
\makecell{98.90\\$\pm$0.02} &
0.58 M \\
FeatureFox (Ours) & 
\makecell{--\\--} &
\makecell{--\\--} &
\makecell{99.91\\$\pm$0.00} &
\makecell{\textbf{99.45}\\$\pm$0.00} &
-- \\
\bottomrule
\end{tabular}
}
\end{table}

\subsection{Feature-level performance on full MFInstSeg training dataset}

We compare feature-level recognition and localization performance of AAGNet and FeatureFox, both trained on MFInstSeg. In contrast to prior work that reports instance grouping and semantic labeling separately, we formulate AFR on B-Rep faces as a single panoptic feature-level graph segmentation task: each face is assigned a feature instance ID and a semantic class and the result can be quantified using Panoptic Quality (PQ) as metric. This unifies instance and semantic segmentation in one output, analogous to panoptic segmentation of image pixels in computer vision.

\paragraph{Recognition \& Localization Accuracy}

We first report Recognition \& Localization Accuracy, defined as the fraction of ground-truth machining feature instances that are fully recovered meaning that all faces of the instance are assigned to a single predicted feature instance and the semantic class being predicted correctly. Since AAGNet outputs instance grouping and semantic segmentation separately, we convert per-face semantic predictions into per-instance labels.
Let $\hat{C}\subseteq V$ be a predicted feature instance, and let $\mathbf{s}_v\in\mathbb{R}^{K}$ denote the pre-softmax semantic logits for face $v$ over $K$ machining feature classes.
We perform logit-sum voting by aggregating
\[
\mathbf{S}_{\hat{C}}=\sum_{v\in \hat{C}} \mathbf{s}_v,
\]
and assign the instance class as
\[
\hat{c}_{\hat{C}}=\arg\max_{k\in\{1,\dots,K\}} \left(\mathbf{S}_{\hat{C}}\right)_k.
\]
Equivalently, this corresponds to maximizing the product of per-face class posteriors under a conditional independence assumption.

\paragraph{Panoptic Quality (PQ) for machining features.}
Again, let ground-truth machining features be instances $\mathcal{G}=\{(C, c)\}$ where
$C\subseteq V$ is the set of faces belonging to one feature instance and $c$ is its machining feature class.
Similarly, predictions are $\mathcal{P}=\{(\hat{C}, \hat{c})\}$.
For a predicted instance $\hat{C}$ and a ground-truth instance $C$, define the face-based IoU as
\[
\mathrm{IoU}(\hat{C},C)=\frac{|\hat{C}\cap C|}{|\hat{C}\cup C|}.
\]
We form a one-to-one matching $\mathrm{TP}\subseteq \mathcal{P}\times \mathcal{G}$ over pairs with
$\hat{c}=c$ and $\mathrm{IoU}(\hat{C},C)>0.5$ (unmatched predictions are $\mathrm{FP}$ and unmatched ground truth are $\mathrm{FN}$).
Then
\[
\mathrm{PQ}
=\frac{\sum\limits_{((\hat{C},\hat{c}),(C,c))\in \mathrm{TP}} \mathrm{IoU}(\hat{C},C)}
{|\mathrm{TP}|+\tfrac{1}{2}|\mathrm{FP}|+\tfrac{1}{2}|\mathrm{FN}|}\, .
\]

We believe that PQ is the best metric for evaluating the AFR task as it scores the joint correctness of "what machining feature is it?" and "which exact faces belong to that feature?", in one number. With this metric we couple recognition and localization and penalize merges/splits via FP/FN while rewarding accurate face-level localization via IoU.

For both AAGNet and FeatureFox trained on the full MFInstSeg training set, we report results over three random seeds. For FeatureFox, the \texttt{XGBoost} components were effectively seed-insensitive, yielding identical performance across seeds. We used the exact same hyperparameters as described in the paper for AAGNet. Results are shown on the right side of table \ref{tab:mfinstseg_featurelevel_2sizes}. AAGNet achieves slightly higher PQ results compared to FeatureFox, while FeatureFox has a marginal edge in terms of features recognized completely.

\begin{table}[t]
\centering
\caption{Per-Feature Performance on MFInstSeg at Two Training Set Sizes Excluding Stock Features}
\label{tab:mfinstseg_featurelevel_2sizes}
\renewcommand{\arraystretch}{1.15}
\setlength{\tabcolsep}{6pt}

\resizebox{\columnwidth}{!}{%
\begin{tabular}{lcccc}
\toprule
\multirow{2}{*}{Network} &
\multicolumn{2}{c}{1000 training samples} &
\multicolumn{2}{c}{Full training set (43,524 samples)} \\
\cmidrule(lr){2-3}\cmidrule(lr){4-5}
& \makecell{Recognition \&\\Localization Acc(\%) $\uparrow$}
& \makecell{PQ(\%) $\uparrow$}
& \makecell{Recognition \&\\Localization Acc(\%) $\uparrow$}
& \makecell{PQ(\%) $\uparrow$} \\
\midrule
AAGNet &
\makecell{65.54\\$\pm$2.87} & \makecell{78.81\\$\pm$2.28} &
\makecell{95.49\\$\pm$2.38} & \makecell{\textbf{97.44}\\$\pm$1.16} \\
FeatureFox (Ours) &
\makecell{\textbf{93.43}\\$\pm$0.00} & \makecell{\textbf{94.65}\\$\pm$0.00} &
\makecell{\textbf{95.99}\\$\pm$0.00} & \makecell{96.87\\$\pm$0.00} \\
\bottomrule
\end{tabular}%
}
\end{table}

\subsection{Sample efficiency on partial MFInstSeg training dataset}

\begin{figure}[t]
\centering
\includegraphics[width=\columnwidth]{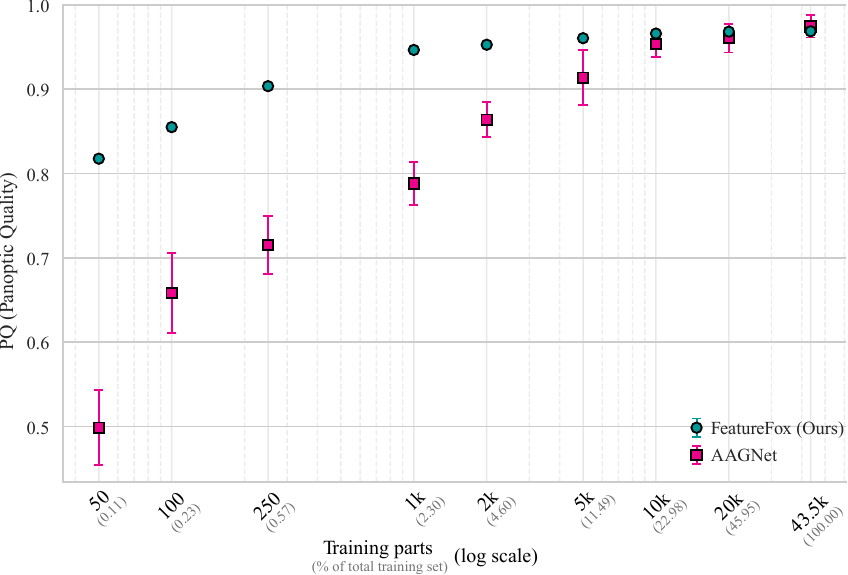}
\caption{Sample efficiency on MFInstSeg (excluding stock): PQ versus training-set size for AAGNet and FeatureFox. FeatureFox dominates in the low-data regime and reaches $>90\%$ PQ with $\sim250$ training parts, whereas AAGNet requires $\sim5{,}000$; AAGNet is only marginally better on the full training set.}

\label{fig:featfox_pq_wo_stock_comparison}
\end{figure}

FeatureFox is motivated by the scarcity of large, labeled industrial 3D-CAD datasets. With FeatureFox, we therefore design a model that can recognize machining features reliably with limited training parts. To assess sample efficiency, we train AAGNet and FeatureFox on matched training subsets of MFInstSeg with varying sizes (three random seeds each) and evaluate on the full MFInstSeg test set. For AAGNet, we use the default hyperparameters and apply early stopping since convergence requires more epochs when fewer training samples are available. We also experimented with larger learning rates for smaller training subsets but achieved similar results.  Figure~\ref{fig:featfox_pq_wo_stock_comparison} reports PQ (excluding stock features) for training subsets ranging from $50$ to $43.5$k samples. FeatureFox consistently performs better in the low-data regime (e.g., $0.83$ vs.\ $0.50$ PQ at $50$ samples) and remains ahead up to roughly half of the full training set. AAGNet only slightly surpasses FeatureFox on the full dataset (about +1 PQ point). To exceed $90\%$ PQ, AAGNet requires $\sim5{,}000$ training parts, whereas FeatureFox reaches the same level with $\sim250$. Table~\ref{tab:mfinstseg_featurelevel_2sizes} summarizes results for $1{,}000$ samples and the full training set.

\subsection{Computational efficiency}

We trained both FeatureFox and AAGNet on an NVIDIA RTX 6000 Ada GPU. Training AAGNet (excluding data loading) for 100 epochs on the whole dataset took $38$ minutes with an average GPU utilization of $18\%$. We did not adapt AAGNet for optimized GPU utilization. Training FeatureFox on the whole dataset and the same machine took $8$ seconds, with $100\%$ GPU utilization.

\subsection{Ablation study on industrially complex CAD parts}

\begin{figure}[t]
\centering
\includegraphics[width=0.7\columnwidth]{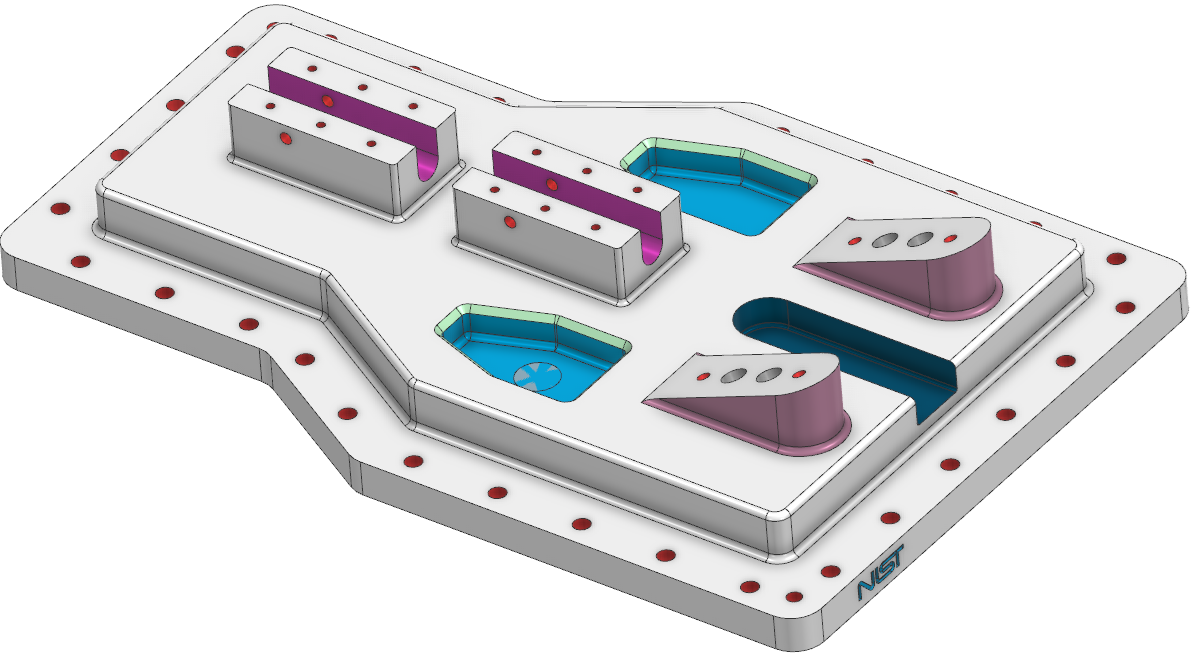}
\caption{Inference on an industrial CAD part from the NIST dataset \cite{lipman_nist_cad_pmi}. Most features are recognized correctly. Chamfers were labeled and correctly recognized; edge blends were not labeled. The dark grey multi-step holes on the inclined plane are correctly grouped as instances but misclassified due to very few multi-step-hole training samples. The two pink vertical features around the inclined planes are predicted as freeform but should be stock. The \textit{NIST} logo is segmented into two pocket instances: \textit{N} and \textit{IST} (bottom of figure).}
\label{fig:inference_nist}
\end{figure}

To the best of our knowledge, none of the current State-of-the-Art models has been tested on industrial CAD data. The sample efficiency of our model makes it possible to label and train FeatureFox on a small dataset of $270$ industrially realistic NX .prt files. We demonstrate the panoptic segmentation capabilities on one of the publicly available National Institute of Standards and Technology (NIST) 3D-CAD files \cite{lipman_nist_cad_pmi}, which we did not train on. The result is shown in figure \ref{fig:inference_nist}. Holes, pockets and slots were all recognized correctly. Minor prediction errors exist, but most features were recognized reliably.

\section{Discussion and Limitations}

FeatureFox shows that panoptic AFR can be achieved with a hierarchical tree-based pipeline: a calibrated binary edge-boundary model yields stable feature instances, and a per-instance classifier assigns semantic labels from aggregated instance attributes. This design excels in the low-data regime, reaching high PQ with a few hundred training parts, whereas deep AFR architectures typically require thousands. On the full MFInstSeg training set, AAGNet slightly surpasses FeatureFox in PQ, suggesting end-to-end representation learning can exploit subtle geometric cues when sufficient data is available. We also advocate Panoptic Quality (PQ) as a unified metric that jointly captures instance separation and semantic correctness.

A key limitation of FeatureFox is its non-end-to-end nature: boundary errors propagate to semantic labeling. While boundary prediction is strong on MFInstSeg, more ambiguous boundaries would reduce robustness. More broadly, we believe future progress in AFR to require labeled industrial CAD data beyond synthetic data, since real parts contain multi-operation and intersecting features whose instances may be disconnected face sets in the B-Rep yet must be treated as a single manufacturable feature - our method does not handle such cases. Finally, deployment remains unclear (cloud or on-premise), but FeatureFox’s low compute and data requirements make customer-specific on-premise training and inference feasible, which can be beneficial for privacy-sensitive industrial workflows.

Furthermore, we hypothesize that FeatureFox can generalize to variable-sized CAD graphs at inference time, unlike many GNN-based approaches that are sensitive to shifts in graph structure, as long as edge-boundary and instance-subgraph statistics remain comparable across domains. Moreover, inference is linear in graph size, i.e., $\mathcal{O}(|V|+|E|)$. A systematic cross-distribution evaluation is left for future work.

\section{Conclusion}

With FeatureFox, we introduce a panoptic graph segmentation approach for AFR that is competitive with current SOTA methods such as AAGNet~\cite{aagnet_wu2024aagnet}. Unlike prior work that evaluates instance grouping and semantic segmentation separately, we frame AFR as a single joint task and advocate Panoptic Quality (PQ) as the most faithful metric for feature-level recognition and localization. On MFInstSeg, FeatureFox requires only a fraction of the training time and is markedly more sample-efficient than AAGNet: for training sets below $10{,}000$ parts it consistently achieves higher PQ, reaching $\mathrm{PQ}>0.9$ with only $250$ training CAD models, whereas AAGNet requires about $5{,}000$. Finally, to demonstrate industrial applicability, we train on $270$ manually labeled real-world CAD parts and show on a National Institute of Standards and Technology test sample that FeatureFox successfully performs AFR on industrial 3D-CAD geometry.

\section*{Acknowledgment}

Acknowledgments will be added after review due to the anonymized submission format.

\bibliographystyle{splncs04}
\bibliography{mybibliography}

\end{document}